\documentclass[epj]{svjour}

\pdfoutput=1
\usepackage{graphicx}
\usepackage{fancyhdr}
\def\makeheadbox{{
 }}
\def\mytitle{My title} 
\def\myauthors{My name}  
\def\mytype{My type of session}
\def\mysession{My session}

\def\mytitle{The anomaly in the cosmic-ray positron spectrum} 
\def\myauthors{C. H. Chung, H. Gast, J. Olzem and S. Schael}    
\def\mytype{Contributed Talk}    
\def\mysession{Cosmology and Astrophysics}

\newcommand{\+}{$\,$}
\newcommand{\AMS}{\mbox{AMS-01}}
\def\GeV{\ifmmode {\mathrm{\ Ge\kern -0.1em V}}\else\textrm{Ge\kern -0.1em V}\fi}%
\def\TeV{\ifmmode {\mathrm{\ Te\kern -0.1em V}}\else\textrm{Te\kern -0.1em V}\fi}%

\begin{document}
\title{The anomaly in the cosmic-ray positron spectrum}
\author{C. H. Chung \and H. Gast \and J. Olzem\thanks{\emph{E-mail:} jan.olzem@cern.ch} \and S. Schael}
%
%
\institute{1. Physikalisches Institut B, RWTH Aachen University, Germany}
%
\date{}
\abstract{ 
  A recent analysis of cosmic-ray data from a space borne experiment
  by the AMS collaboration supports the observation of an excess in
  the cosmic-ray positron spectrum by previous balloon
  experiments. The combination of the various experimental data
  establishes a deviation from the expected background with a
  significance of more than four standard deviations. The observed
  change in the spectral index cannot be explained without introducing
  a new source of positrons. When interpreted within the MSSM a
  consistent description of the antiproton spectrum, the diffuse
  gamma-ray flux and the positron fraction is obtained which is
  compatible with all other experimental data, including recent WMAP
  data.
} 
\maketitle

\section{Introduction}\label{intro}
Among the cosmic-ray species, antiparticles and diffuse $\gamma$-rays
are of particular interest because they are produced secondarily in
hadronic interactions of protons and nuclei with the interstellar
medium at low rates. Their small abundance makes them a sensitive
probe for the existence of additional -- and possibly exotic --
cosmic-ray sources which would be visible as an excess of particles
above conventional expectations.

One of the most important unsolved questions in modern cosmology is
the nature of dark matter. The most promising dark matter candidate is
the weakly interacting lightest neutralino, $\chi_1^0$, predicted by
supersymmetric extensions to the standard model of particle physics.
The annihilation of neutralinos might constitute an additional primary
source of particles with a unique spectral shape which would be
determined by the parameters of supersymmetry, allowing to put
constraints on new physics beyond the standard model.

A recent reanalysis of the data from the \AMS{}
spectrometer~\cite{aguilar07a} supports the observation of an excess
of cosmic-ray positrons by the HEAT experiments~\cite{beatty04a}. In
this work, we discuss the combined results on the cosmic-ray positron
fraction $e^+ / (e^+ + e^-)$. Assuming that dark matter is largely
constituted by neutralinos, we determine the cosmic-ray preferred
parameter space of the minimal supersymmetric standard model (MSSM)
from a simultaneous fit to the cosmic-ray positron, antiproton and
diffuse $\gamma$-ray data.

\section{Cosmic-ray particle propagation}\label{sec:models} The public
GALPROP code~\cite{strong98a} has been used to model cosmic-ray
particle propagation and calculate the particle spectra as observed
near Earth. GALPROP solves the propagation equation in a diffusion
model with a given source distribution for all cosmic-ray species
and includes convection, diffuse reacceleration, energy loss,
fragmentation and decay in the interstellar medium. The injection
spectra of nuclei and electrons before propagation are assumed to be
power laws in momentum, and their spectral indices, $\gamma_s$ and
$\gamma_e$, respectively, are chosen such that the model reproduces
the most recent cosmic-ray flux measurements.

From fitting the propagation model to electron and proton flux data we
find the most probable values of these indices to be $\gamma_s =
2.35\pm 0.03$ and $\gamma_e = 2.50\pm 0.04$. In order to determine
their errors and thus estimate the uncertainties of the model
predictions, the indices have been varied over small intervals around
their most probable values and the resulting predicted fluxes have
been compared to the data. The $\chi^2$ calculated from the deviation
of the data from the respective prediction gives the 1$\sigma$ errors
of the injection spectral indices.

\begin{figure}[htb]
\begin{center}
  \includegraphics[width=7cm]{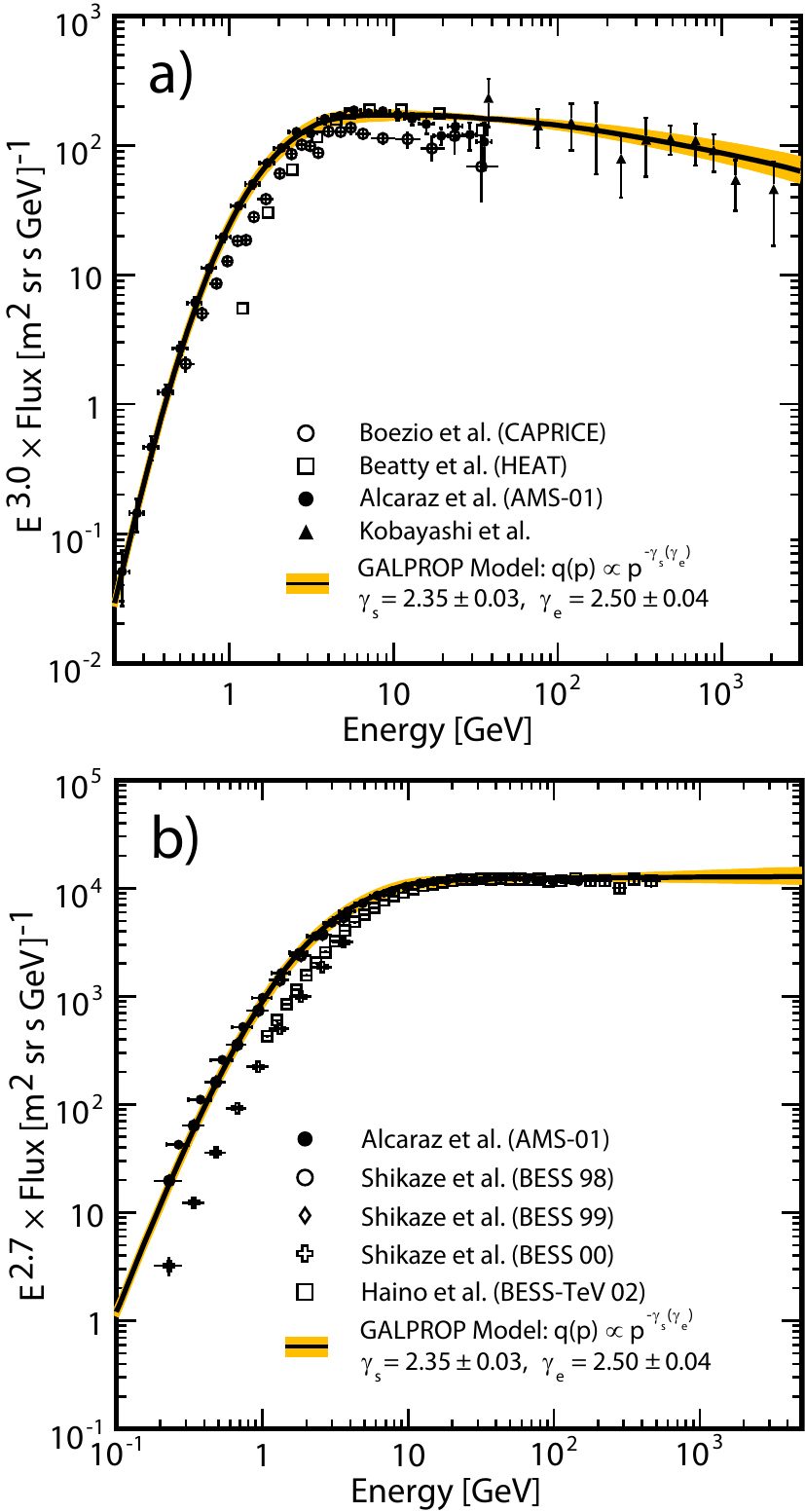}
  \caption{a) Compilation of cosmic-ray electron flux data from
    CAPRICE~\cite{boezio00a}, HEAT-e$^{\pm}$~\cite{beatty04a},
    \AMS{}~\cite{alcaraz00a} and Kobayashi et al.~\cite{kobayashi99a}.
    b) Proton flux data from \AMS{}~\cite{alcaraz00c} and
    BESS~\cite{shikaze07a,haino04a}. In both panels, the solid line
    denotes our GALPROP model and its uncertainty.}
\label{fig:spectra}
\end{center}
\end{figure}

Fig.~\ref{fig:spectra} shows the calculated fluxes of electrons and
protons which are in excellent agreement with the experimental data
over large energy intervals. The uncertainties of the propagation
model -- the fluxes calculated with the injection indices at their
error limits -- are denoted by the yellow areas.  Below energies of
several \GeV{}, the individual measurements differ from each other due
to the time-dependent effect of solar modulation.

\section{The cosmic-ray positron fraction}\label{sec:positrons} 
\begin{figure}[!t]
\begin{center}
  \includegraphics[width=7.9cm]{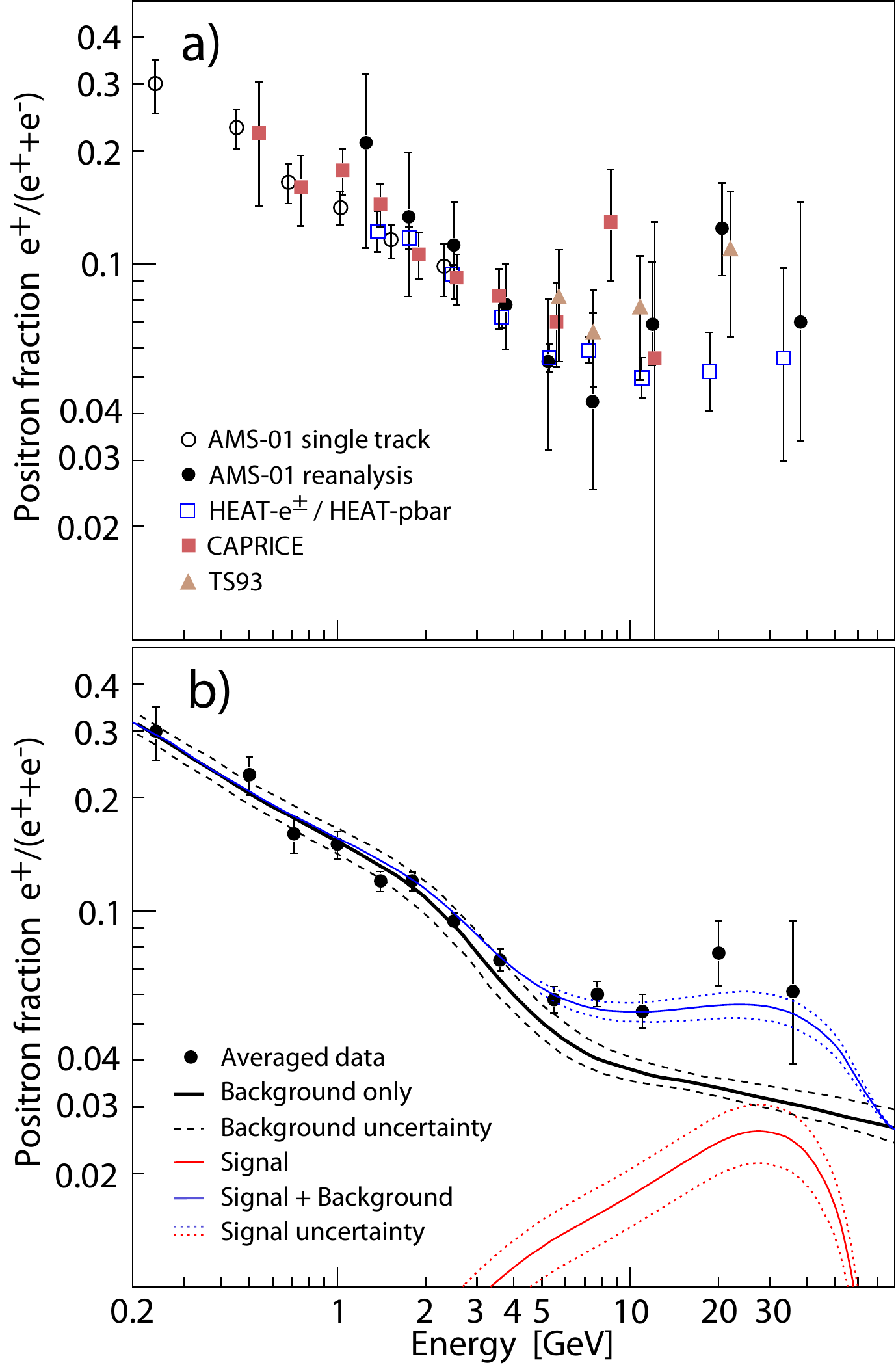}
  \caption{a) Compilation of recent cosmic-ray positron fraction data
    from: \AMS{} (2000)~\cite{alcaraz00b}, the \AMS{}
    reanalysis~\cite{aguilar07a}, HEAT-e$^{\pm}$ and
    HEAT-pbar~\cite{beatty04a}, CAPRICE~\cite{boezio00a} and
    TS93~\cite{golden96a}. b) The combined data together with the
    background model (thick solid line) and its uncertainty (dashed
    lines) as well as the neutralino annihilation signal and
    signal+background for the best fit parameter set (thin solid
    lines). The dotted lines denote the propagation uncertainties of
    the signal contribution.}
\label{fig:positrons}
\end{center}
\end{figure}

The challenge of cosmic-ray positron measurements is the rejection of
the vast proton background. A number of balloon borne experiments have
delivered positron flux data in the energy range from 0.5 to 50\+\GeV,
such as HEAT-e$^{\pm}$ and HEAT-pbar~\cite{beatty04a},
CAPRICE~\cite{boezio00a} and TS93~\cite{golden96a}. Additionally, the
\AMS{} spectrometer has measured the positron flux up to
3\+\GeV~\cite{alcaraz00b} in a low Earth orbit. In order to extend the
sensitivity of \AMS{} to energies of up to 50\+\GeV{}, a reanalysis of
the data has been conducted~\cite{aguilar07a} using the conversion of
bremsstrahlung photons from positrons to achieve a proton background
suppression of more than $10^5$. The result is shown in panel a) of
Fig.~\ref{fig:positrons} together with previous data.

In order to simplify data handling, the measurements on the positron
fraction displayed in Fig.~\ref{fig:positrons} a) have been combined
into one single data set with regard to asymmetric statistical and
systematic errors. Details of this procedure as well as a result table
are given in Ref.~\cite{olzem07a}. Panel b) of
Fig.~\ref{fig:positrons} shows the combined data together with the
model prediction. Above energies of 6\+\GeV{}, the data exhibit a
change in the spectral index of positrons which is clearly
incompatible with the expectation for purely secondary positron
production. Taking into account experimental errors as well as the
model uncertainty, the significance of the deviation amounts to more
than four standard deviations. There is no set of propagation
parameters based on which the GALPROP model would match the data
satisfactorily.  Consequently, the excess in the positron flux cannot
be explained by the current propagation models and thus requires a new
primary source of positrons.

\section{The spectra of cosmic-ray antiprotons and diffuse $\gamma$-rays}\label{sec:antiprotonsAndGammas}
\begin{figure}[!t]
\begin{center}
  \includegraphics[width=7.9cm]{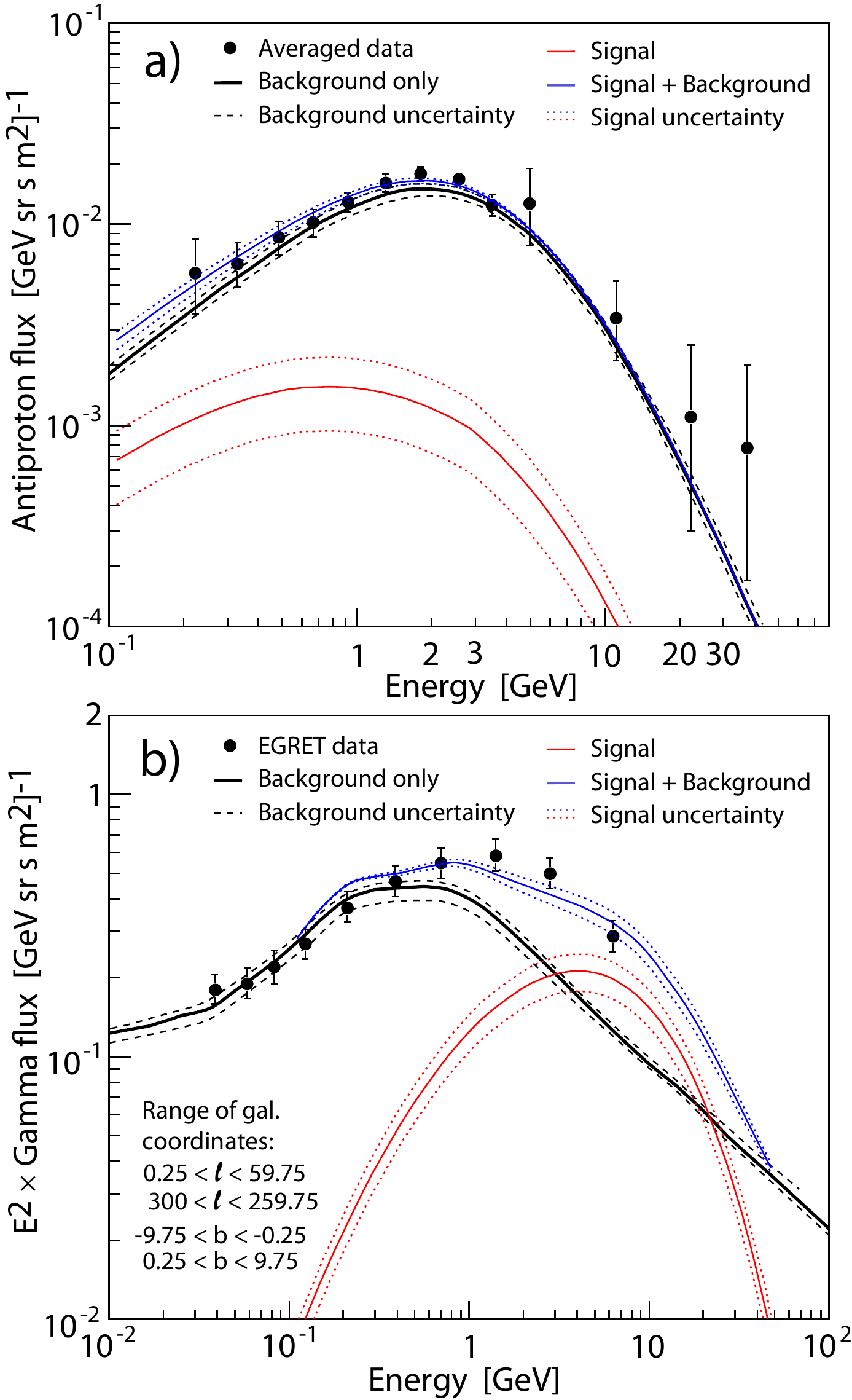}
  \caption{a) The combined antiproton flux data. b) The diffuse
    $\gamma$-ray flux data from EGRET~\cite{hunter97a,chung07a}. Both
    panels: modeled background (thick solid line) and its uncertainty
    (dashed lines) as well as the neutralino annihilation signal and
    signal+background for the best fit parameter set (thin solid
    lines). The dotted lines denote the propagation uncertainties of
    the signal contribution.}
\label{fig:antiprotonsAndGammas}
\end{center}
\end{figure}
Using the same procedure as stated in \S~\ref{sec:positrons},
measurements of the cosmic-ray antiproton flux from
\AMS{} \cite{aguilar02a}, BESS97~\cite{orito00a},
BESS00~\cite{asaoka02a} and CAPRICE~\cite{boezio01a} have been
combined into one single data set. The result is displayed in panel a)
of Fig.~\ref{fig:antiprotonsAndGammas} together with the GALPROP
calculation. Within the experimental errors, the combined data are well in
agreement with the expectation for purely secondary antiproton
production.

Fig.~\ref{fig:antiprotonsAndGammas} b) shows the flux of diffuse
$\gamma$-rays measured by the EGRET~\cite{hunter97a,chung07a}
experiment. Above energies of 1\+\GeV{}, the data exhibit a
significant excess with respect to the model calculation for purely
secondary production which has been interpreted as an additional
primary source of $\gamma$-rays from neutralino
annihilations~\cite{deboer06a}. However, the particular model was
claimed to be in conflict with the observed antiproton
spectrum~\cite{bergstrom06a}. It has recently been pointed out that
the excess could also be an artifact from energy miscalibration of the
experiment~\cite{stecker07a}. The discrepancy is in principle
resolvable by fine-tuning the propagation model parameters, which
however results in predictions for other particle species' spectra
which are incompatible with experimental data~\cite{strong07a}.

\section{Interpretation of the cosmic-ray spectra within the MSSM}\label{sec:MSSM}
\subsection{Constraints on the MSSM parameter space}\label{subsec:constraints} 
Measurements of several quantities are used to constrain the parameter
space of the MSSM, such as the dark matter relic density from
WMAP~\cite{spergel06a} and the branching ratios of the rare decays
$b\rightarrow s\gamma$~\cite{barberio06a} or
$B_S\rightarrow\mu\mu$~\cite{yao06a}. Additional constraints come from
the LEP2 experiments as lower limits on the
neutralio~\cite{heister04a} and neutral Higgs boson
masses~\cite{lep06a}. Furthermore, measurements of the anomalous
magnetic moment of the muon~\cite{bennett06a} suggest low values of
the MSSM parameters $m_0$ and $m_{1/2}$. Fig.~\ref{fig:m0m12} shows
the plane spanned by $m_0$ and $m_{1/2}$ for $\tan\beta=40$, $A_0=0$
and sign$\,\mu=+1$ together with the respective $2\sigma$ limits
derived from the above constraints.

\subsection{MSSM parameter scan with cosmic-ray data}\label{subsec:scan} 
In order to put further constraints on the MSSM parameter space from
cosmic-ray data, we have conducted scans of the plane spanned by the
parameters $m_0$ and $m_{1/2}$ for particular fixed values of
$\tan\beta$. For each of the sample points in the plane the
contributions to the positron fraction and the antiproton and
$\gamma$-ray spectra from neutralino annihilation after galactic
propagation have been calculated and simultaneously fitted to the
experimental data together with the GALPROP models for the purely
secondary background components. For the calculations, an isothermal
dark matter halo profile with a local density of $\rho_0 =
0.3\+\GeV{}/$cm$^3$ has been assumed. They were performed using the
public DarkSUSY 4.1~\cite{gondolo04a}, FeynHiggs
1.2.2~\cite{heinemeyer00a} and ISAJET 7.75~\cite{baer03a} packages
with the top-quark mass fixed to $m_{\mathrm{top}}$=170.9\+\GeV{}, $A_0=0$ and
sign\+$\mu=+1$.
\begin{figure}[!t]
\begin{center}
  \includegraphics[width=8cm]{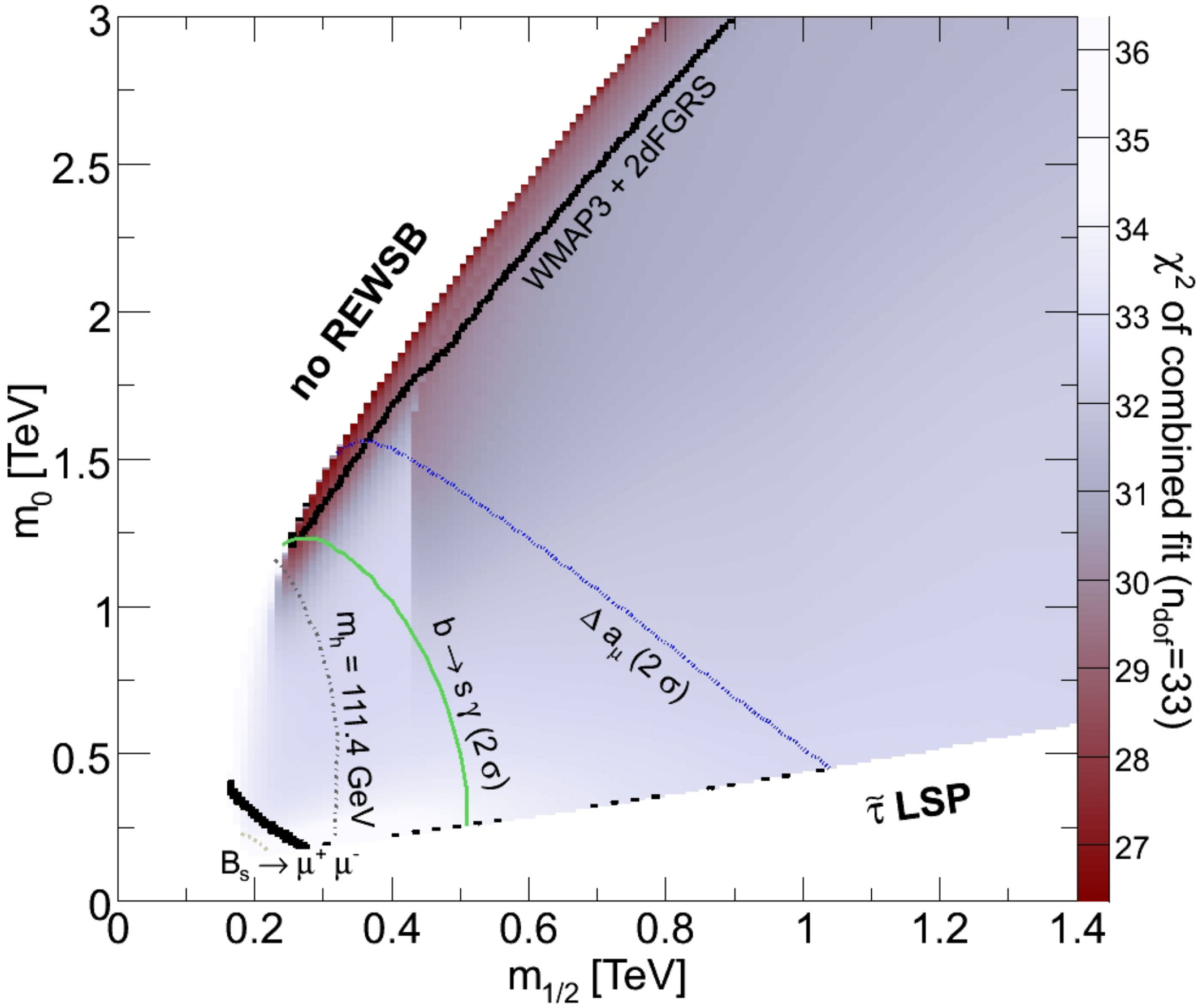}
  \caption{The plane spanned by the MSSM parameters $m_0$ and
    $m_{1/2}$ for $\tan\beta=40$, $A_0=0$ and sign$\,\mu=+1$. Current
    experimental constraints ($2\sigma$) are denoted by the solid,
    dotted and dash-dotted lines. The color scale gives the
    $\chi^2$ of the MSSM fit to the cosmic-ray data.}
\label{fig:m0m12}
\end{center}
\end{figure}

The particle fluxes $\Phi_i$ as observed near Earth can be described
by adding the calculated signal contributions, $S_i$, for the
individual particle species $i$ to the respective GALPROP background
models, $B_i$, according to $\Phi_i = B_i + f_i\cdot S_i$. Here, the
$f_i$ denote individual boost factors to allow for a signal
enhancement due to a possible clumpy nature of the dark matter
distribution in the solar neighborhood. In this case, we expect the
individual boost factors to differ significantly from each other due
to the different travel paths of the particle species which are
determined by their mean energy loss. In particular, the boost factor
for the antiproton signal should be small with respect to the others,
since the low synchrotron radiation level of heavy particles allows
them to be measured almost independently from their production
location in the galaxy. The boost factors were determined as free
parameters in fits of the $\Phi_i$ to the experimental data described
in \S\+\ref{sec:positrons} and \S\+\ref{sec:antiprotonsAndGammas}.

Fig.~\ref{fig:m0m12} shows the combined $\chi^2$ from the simultaneous
fits as a function of $m_0$ and $m_{1/2}$ for $\tan\beta=40$.
Apparently, the cosmic-ray data clearly favor the focus point region
at large values of $m_0$, and we find the best fit parameters to be
$m_0=1230\+\GeV{}$ and $m_{1/2}=260\+\GeV{}$. This point is well in
agreement with all constraints on the MSSM parameter space stated in
\S\+\ref{subsec:constraints}, including recent data from WMAP.

The contributions from neutralino annihilation to the individual
particle spectra as well as the signal + background curves for the
best fit parameter set are shown in Fig.~\ref{fig:positrons} b)
and~\ref{fig:antiprotonsAndGammas} in comparison with the experimental
data. With the additional primary cosmic-ray component from neutralino
annihilation, the experimental data for the positron fraction and the
spectra of antiprotons and $\gamma$-rays can well be reproduced. The
combined $\chi^2$ turns out to be 28 with 33 degrees of freedom, and
the boost factors $f_i$ are found to be $85\pm 15$ for positrons,
$1\pm 0.5$ in the case of antiprotons and $310\pm 50$ for
$\gamma$-rays.

In the region of the parameter space preferred by the cosmic-ray data,
neutralinos have a significant higgsino component of more than 30\+\%
and dominantly annihilate into W-boson pairs via t-channel exchange of
charginos. For the best fit parameters, we find the mass of the
$\chi_1^0$ to be 91\+\GeV{} and a value for the mass of the lightest
Higgs boson of 113.7\+\GeV{}.

\subsection{Dependence on $\tan\beta$ and $m_{\mathrm{top}}$}\label{subsec:nuisance}
The choice of $\tan\beta$ is critical to constrain the MSSM parameter
space with cosmic-ray data. For varying values of $\tan\beta$, the
combined fits favor a neutralino mass between 80\+\GeV{} and
120\+\GeV{}. Unless $\tan\beta$ is higher than 50, we always find an
overlap of the parameter space favored by cosmic rays with the relic
density constraints from WMAP in the focus point region. Furthermore,
the preference of cosmic rays in terms of the MSSM parameter space is
sensitive to the mass of the top-quark, whose value is currently known
with a precision of 1.8\+\GeV{}\cite{TEWG}. In particular, for low
$\tan\beta$ and values of $m_{\mathrm{top}}>173\+\GeV{}$, the focus
point region is not available unless $m_0$ is larger than about
3\+\TeV{}. In order to put accurate constraints on the MSSM parameter
space from cosmic-ray data, the impact of varying values of
$\tan\beta$ and $m_{\mathrm{top}}$ has to be investigated further.

\section{Conclusions}\label{sec:conclusions}
In this work, the combined recent experimental results on the
cosmic-ray positron fraction have been presented. The data exhibit an
excess of positrons above energies of 6\+\GeV{} which cannot be
explained by purely secondary positron production alone and thus
requires an additional primary source of positrons. In this work, we
interpret this source to be the annihilation of supersymmetric
neutralinos constituting dark matter. A simultaneous fit to the
cosmic-ray positron, antiproton and $\gamma$-ray data shows that, for
particular sets of the MSSM parameters, this hypothesis gives a fully
consistent description of the cosmic-ray spectra which is compatible
with all other experimental data. We find that the cosmic-ray data
clearly prefer the focus point region of the MSSM parameter space but
reveal almost no sensitivity to $\tan\beta$.

\end{document}